\documentclass[3pt,onecolumn]{article}
\usepackage{graphicx}

\newcommand{\ben}{\begin{enumerate}}
\newcommand{\een}{\end{enumerate}}

\newcommand{\be}{\begin{equation}}
\newcommand{\ee}{\end{equation}}
\newcommand{\bea}{\begin{eqnarray}}
\newcommand{\eea}{\end{eqnarray}}

\begin{document}
\begin{flushright}
\end{flushright}
\vspace{0.1cm}
\thispagestyle{empty}

\begin{center}
{\Large\bf 
Pair-correlation analysis of HD~10180 reveals a possible planetary orbit at about 0.92 AU
}\\[13mm]
{\rm  Kasper Olsen\footnote{The Department of Physics, Technical University of Denmark, 2800 Kongens Lyngby, Denmark.  E-mail: kasper.olsen@fysik.dtu.dk} and Jakob Bohr\footnote{Fuel Cell and Solid State Chemistry Division, Ris{\o} DTU, Technical University of Denmark, 4000 Roskilde, Denmark. E-mail: jabo@risoe.dtu.dk}}\\[18mm]

\end{center}

\noindent Since the observation in the first half of the nineties of exoplanets by precise timing measurements of a microsecond pulsar \cite{wolszczan1992} as well as the observation of a Jupiter-like exoplanet around a main-sequence star \cite{mayor1995} around 500 exoplanets have been observed by techniques such as microlensing, astrometry, transiting, and imaging \cite{schneider2010}.  Relatively recently, exoplanets with near Earth-like orbit have been reported  \cite{beaulieu2006} as well as planets with near Earth-like size \cite{Mayor2009}. The surge in exoplanet research and their observation has also given renewed impetus to discussions of planetary system formation \cite{barnes2010} and to the century-old discussion \cite{nieto1972} of possible distance laws or regularities \cite{lovis2010,young2010,poveda2008}.

Analyses for distance regularities akin to the Titius-Bode rule have suffered from what at times can be seen as arbitrary numbering schemes for planets
(for a discussion of the undesirable in having to introduce additional parameters or other degrees of freedom, see 
Ref. \cite{lovis2010}). 
We have suggested a pair-correlation analysis that does not involve any numbering schemes for the planets \cite{bohr2010}.  Further, an analysis of the Solar System demonstrated the correlation between the orbits of the planets extending far beyond nearest neighbour orbits. In this letter, we analyze the pair-correlations between the logarithmic positions of the exoplanets in HD 10180.
This exoplanetary system has six planets (plus one unconfirmed) surrounding a sunlike star  \cite{lovis2010}.
The results of the analysis show that the six planets have a pair-correlation with nearly equally spaced peaks revealing the regularity of this system and its long-ranged correlation.  Interestingly, the pair-correlation of the six planets reveals that a seventh orbit consistent with this group of six planets exists. There is no indication for an eighth orbit. A modeling of the pair-correlations shows this additional consistent orbit to be at 0.92$\pm 0.05$ AU.

The correlations between the positions of the exoplanets are determined from the pair-correlation function, which reads
\be
P(\Delta) = \int \rho(x+\Delta)\rho(x)dx.
\ee
Here $\rho(x)$ is a distribution that measures the  positions, $x_i$, of the planets on a logarithmic scale. For $\rho$ we use a sum of Gaussian distributions so that $\rho(x) = \sum\alpha_i \exp^{-(x-x_i)^2/2\sigma^2}$, where $x_i$ is the logarithmic position of planet $i$, $x_i=\log(a_i/10^6$km), $a_i$ is the semi-major axis, and $\sigma =0.075$ corresponding to a full-width-at-half-maximum $\omega =0.18$. All weights, $\alpha_i$, are chosen as unity, $\alpha_i=1$, as the pair-correlation does not depend much on the choice of $\alpha_i$ \cite{bohr2010}. 

The function $P(\Delta)$ is symmetric around the central self-correlation peak at $\Delta=0$. For the general case involving $N$ planets there will be $N(N-1)/2$ peaks to either side of the self-correlation peak. With the Gaussian smearing above this will lead to broad wings. However, if the peaks are equidistantly spaced then the $N(N-1)/2$ peaks will coalesce into a smaller number of peaks.  For the case of six equidistant planets there will be five side-peaks indicating probable distances for the nearest neighbour, next nearest 
neighbour etc.

In Fig. 1a is shown $\rho(x)$ for HD 10180, i.e. the sum of the six corresponding Gaussian functions, and in Fig. 1b is shown the corresponding pair-correlation function, $P(\Delta)$. Fig. 1b display a well structured set of nearly equidistant peaks. 
This is a manifestation of positional correlations in an exoplanetary system.
Interestingly, there are six peaks to either side and not the expected five considering that the curve is generated using six planet positions. This suggests a seventh position. 

A first guess for the seventh position can be found from an inspection of the logarithmic positions of the exoplanets in Fig. 1a. Two candidate positions reveal themselves -- between planets HD~10180{\it~f} and {\it g}, and between planets {\it g} and {\it h}, see Table 1. Fig. 2 corresponds to the first choice and Fig. 3 to the second. Fig. 2b shows enhanced correlations when compared to Fig. 1b, while Fig. 3b shows a significant destruction of correlations when compared with Fig. 1b. We therefore conclude that only a position between HD~10180~{\it f} and {\it g} is consistent with the six observed positions ({\it c}--{\it h}).  By observing the calculated correlation pattern for a mesh of positions between HD~10180~{\it f} and {\it g}, the optimum position of the additional orbit is found to be at $0.92\pm 0.05$ AU.

Recently, an unconfirmed object close to the HD 10180 star has been reported \cite{lovis2010}. When this object is included in the analysis (Fig. 4) the positional correlations are to a large extend destroyed, see Fig. 4b. Therefore, this unconfirmed object is found not to belong to the same group of planets in the sense that its position is not in correlation with the six other reported planets. The seventh object is also significantly closer to the star (0.022~AU).

The strength of the above analysis is that $P(\Delta)$ does not depend on any specific numbering of the planets. For example, it is independent of an interchange of the number index given to the planets. The calculated pair-correlation reveals a seventh possible orbit following a simple inspection of its number of peaks. When the origin of the type of orbit-orbit correlations are better understood we envision that it will be possible to extract quantitative numbers describing the planetary system under analysis and move beyond the qualitative inspections of pair-correlations function reported in this letter.


\vspace{5cm}
\begin{table}[h]
\caption{{\small Planets orbiting the sunlike star HD~10180. Here, $a_i$ is the semimajor axis, mass is in multiples of Jupiter mass, and $x_i$ is $\ln (a/10^6 {\rm km})$. The data for $a_i$ and mass are from \cite{lovis2010}. The two bottom rows are discussed in the text.}}
\label{tab:1}       
\begin{tabular}{lccrrr}
\hline\noalign{\smallskip}
Name &  & $a_i$ / [$10^6$ km] & Mass $/[M_J]$ & $x_i$\\
\noalign{\smallskip}\hline\noalign{\smallskip}
HD 10180  & -- & -- & -- & --   \\
HD 10180 b  & unconfirmed planet & 3.329 & $>$0.00425  & 1.203\\
HD 10180 c  & exoplanet &9.589 &  $>$0.04122  & 2.261\\
HD 10180 d  & exoplanet &19.29 & $>$0.03697  & 2.959\\
HD 10180 e  &  exoplanet &40.38 & $>$0.07897  & 3.698\\
HD 10180 f &  exoplanet &73.74 & $>$0.07520  & 4.301\\
HD 10180 g & exoplanet & 212.7 & $>$0.06733  & 5.360\\
HD 10180 h & exoplanet & 508.6 & $>$0.20262  & 6.232\\
\noalign{\smallskip}\hline
HD 10180 f2  & consistent orbit & 138 & -- & 4.927\\
HD 10180 g2  & inconsistent orbit & 329 & -- & 5.796\\
\noalign{\smallskip}\hline
\end{tabular}
\end{table}
\newpage
\begin{figure}[h]\centering
\caption{(A) Logarithmic position of six planets, not including the unconfirmed HD 10180 {\it b}. (B) Correlation function for these six planets.
Notice that there are six side peaks rather than five. A linear regression of the positions of these six peaks gives an $R^2$ value of 0.996012.}
\vspace{5mm}
\includegraphics[width=5.5cm]{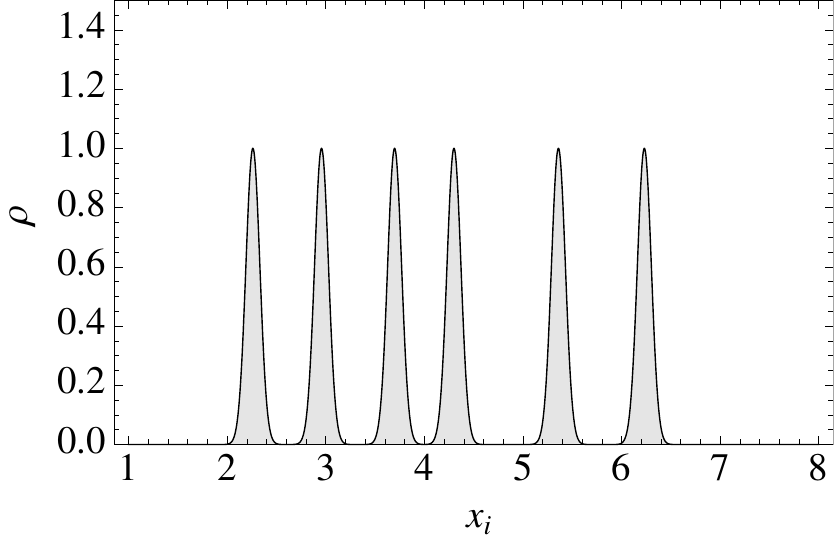}
\includegraphics[width=5.5cm]{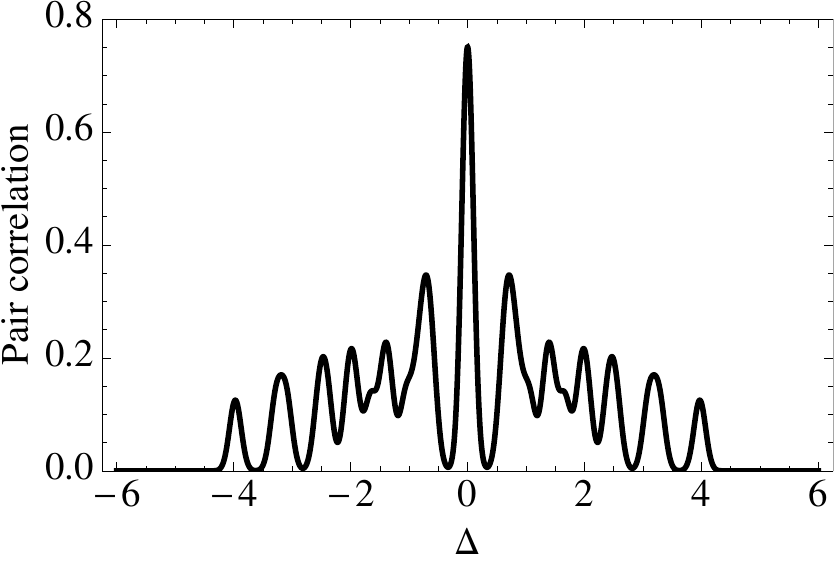}
\end{figure}
\begin{figure}[h]\centering
\caption{(A) Logarithmic position of the six orbits and a seventh at $138 \times 10^6$ km (f2) between {\it f} and {\it g}. (B) Correlation function for these seven orbits. A linear regression of the positions of the six peaks gives an $R^2$ value of 0.99861.}
\vspace{5mm}
\includegraphics[width=5.5cm]{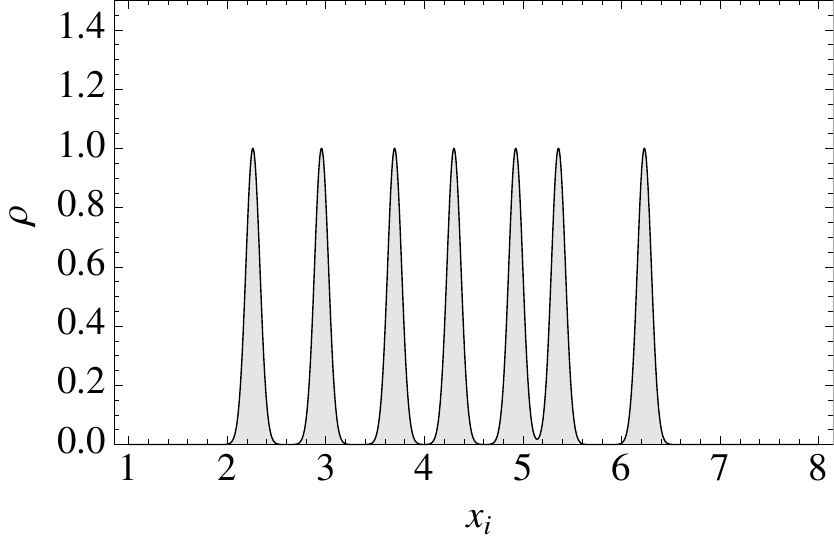}
\includegraphics[width=5.5cm]{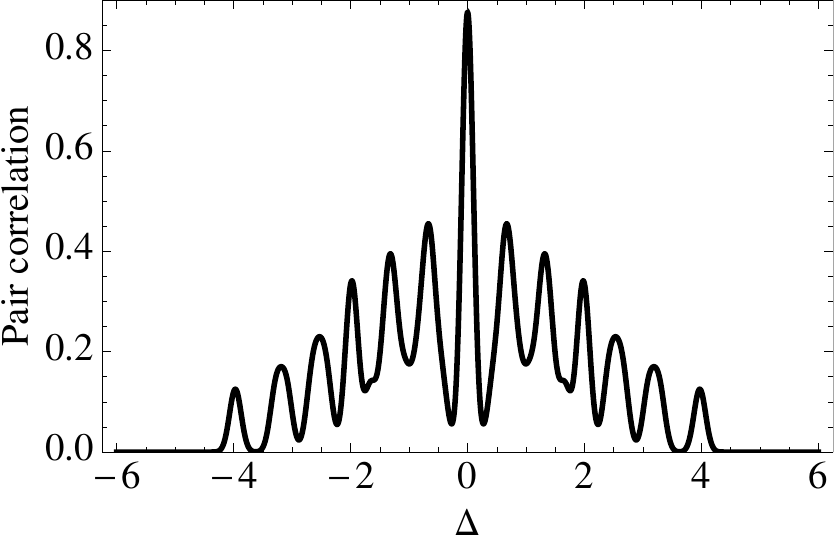}
\end{figure}

\newpage
\begin{figure}[h]\centering
\caption{(A) Logarithmic position of the six orbits and a seventh at $329 \times 10^6$ km (g2) between {\it g} and {\it h}. (B) Correlation function for these seven orbits.}
\vspace{5mm}
\includegraphics[width=5.5cm]{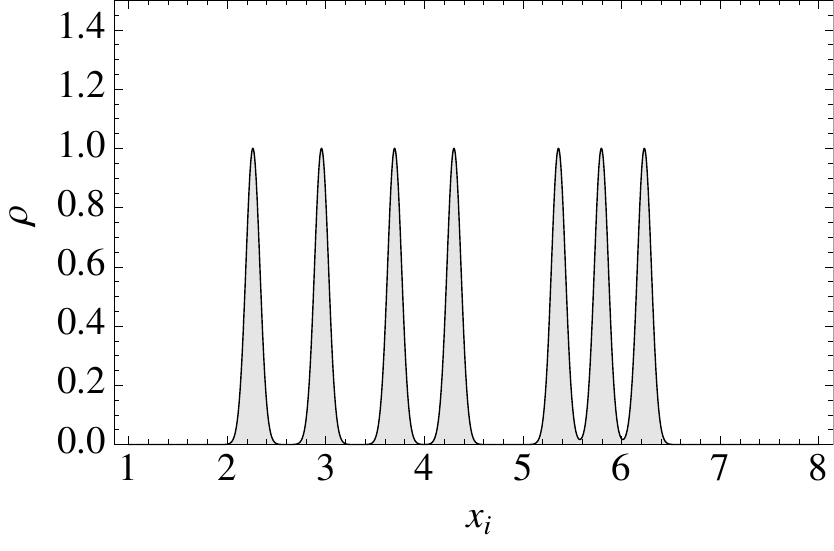}
\includegraphics[width=5.5cm]{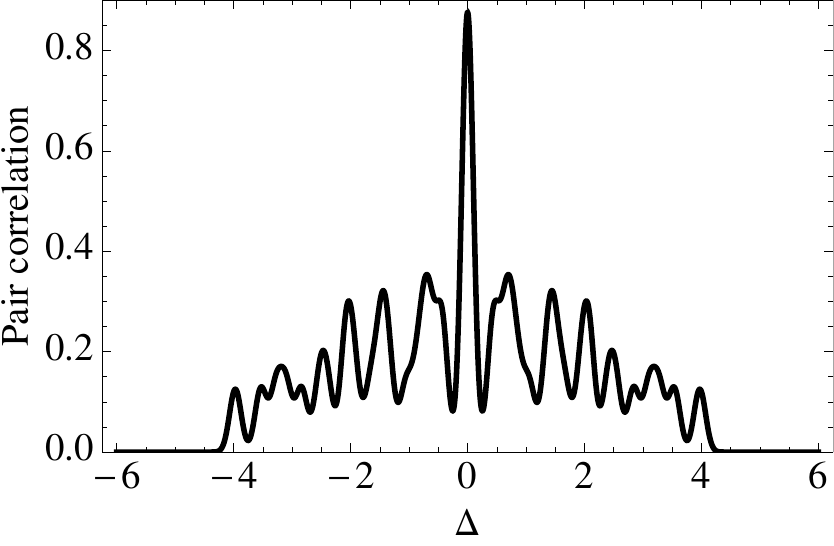}
\end{figure}
\begin{figure}[h]\centering
\caption{(A) Logarithmic position of the six confirmed planets with the HD 10180 {\it b}. (B) Correlation function for these seven planets.}
\vspace{5mm}
\includegraphics[width=5.5cm]{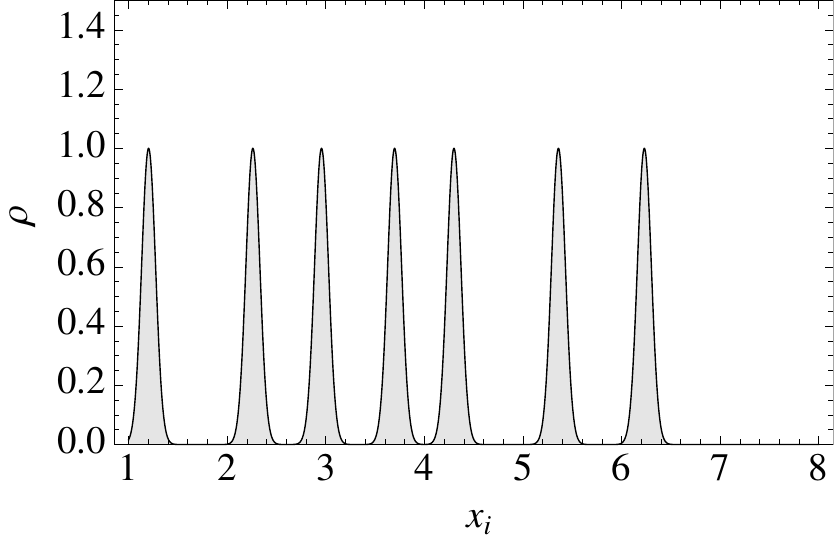}
\includegraphics[width=5.5cm]{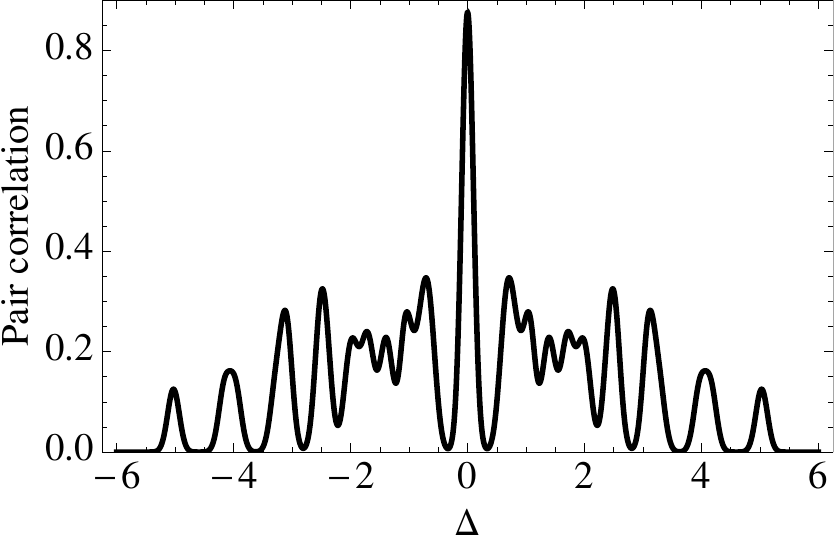}
\end{figure}


\end{document}